\documentclass[amsmath,amssymb,aps,twocolumn,UKenglish,prmaterials]{revtex4-2}
\usepackage{courier}
\usepackage{graphicx}
\usepackage{amsmath,amssymb}
\usepackage{mathrsfs}
\usepackage{bm}
\usepackage{supertabular}
\usepackage{float}
\usepackage{soul}
\usepackage{color}
\usepackage{natbib}
\usepackage{placeins}
\usepackage[version=3]{mhchem}
\usepackage{array}
\usepackage[caption=false]{subfig}
\usepackage{multirow}
\usepackage{isodate}
\usepackage{afterpage}
\usepackage{lipsum}
\usepackage{comment}
\usepackage{color}
\usepackage{titlesec}
\usepackage{soul}

\newcommand{\ie}{i.e., }

\newcommand{\exciting}{{\usefont{T1}{lmtt}{b}{n}exciting}}

\begin{document}

\preprint{APS/123-QED}

\title{Theoretical Insights into Inorganic Antiperovskite Nitrides (X$_3$NA; X = Mg, Sr, Ca, Ba; A = Sb, As): An Emerging Class of Materials for Photovoltaics}

\author{Sanchi Monga$^{*}$}
\author{Manjari Jain}
\affiliation{Department of Physics, Indian Institute of Technology Delhi, New Delhi 110016, India}
\author{Claudia Draxl$^{*}$}
\affiliation{Physics Department and IRIS Adlershof, Humboldt-Universität zu Berlin, 12489 Berlin, Germany}
\author{Saswata Bhattacharya$^{*}$} 
\affiliation{Department of Physics, Indian Institute of Technology Delhi, New Delhi 110016, India}
\email{sanchi@physics.iitd.ac.in[SM], claudia.draxl@physik.hu-berlin.de [CD], saswata@physics.iitd.ac.in [SB]}

\begin{abstract}
\noindent Antiperovskite nitrides are potential candidates for applications harvesting solar light. With a comprehensive state-of-the-art approach combining hybrid density-functional theory, many-body perturbation theory, the Wannier-Mott model, density-functional perturbation theory, and the Feynman polaron model, we explore excitonic and polaronic effects in X$_3$NA (X: Mg, Ca, Sr,  Ba, A = Sb, As). For all of them, we uncover a significant influence of the ionic dielectric screening on the static dielectric constant. Small exciton binding energies, weak electron-phonon coupling, and high charge-carrier mobilities  facilitate enhanced charge transport in Mg$_3$NSb, Sr$_3$NSb, and Ba$_3$NSb. Our results highlight the potential of these nitrides as optimal candidates for efficient photovoltaic absorbers.
\end{abstract}

\maketitle

\begin{figure}
		\centering
		\includegraphics[width=1\columnwidth]{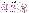}
		\caption{Unit cells of the antiperovskites $\mathrm{X_3NA}$ with space groups (a) $\mathit{Pm\overline{3}m}$, (b) $\mathit{P6_3/mmc}$, and (c) $\mathit{Pbnm}$.}
		\label{structure}
	\end{figure}

Halide perovskites have gained significant attention in the field of photovoltaics due to their exceptional electronic and optical properties, showing a remarkable increase in the photo-conversion efficiency (PCE) from 3.8 $\%$ to 25.5 $\%$  within a decade~\cite{kim2020high,yu2018diversity,nrel}. Despite these advantageous properties, long-term operational instability and the tendency of the devices to degrade over time, make them unsuitable for large-scale commercial applications~\cite{schileo2021lead, ju2018toward}. In contrast, antiperovskites, formed by the electronic inversion of perovskites, are found to be exceptionally stable~\cite{chi2002new,beznosikov2003predicted,gabler2004sr3n,stoiber2019perovskite,heinselman2019thin}. In the resulting structural formula $\mathrm{X_3BA}$, X is a cation and A/B are different-sized anions~\cite{wang2020antiperovskites}. In their ideal cubic structure  shown in Fig.~\ref{structure} on the left, the A-site anion is at the cuboctahedral center and the B-site anion is at the corners surrounded by six X-site cations that form octahedra, BX$_6$. 

The inversion of ions in antiperovskites results in different band-edge characteristics compared to conventional halide perovskites, thus expanding the space of possible compositions. Numerous studies have been conducted on the electronic properties of inorganic antiperovskites, as evidenced in the literature ~\cite{dai2019bi,kang2022antiperovskite,dahbi2022earth,sreedevi2022antiperovskite,liang2022first,kang2023native}. For example, Mochizuki $\mathit{et}$ $\mathit{al}$.~\cite{mochizuki2020theoretical} examining the structural stability and electronic properties of both previously known and newly discovered, but as-yet-unsynthesized, antiperovskites, proposed materials favorable in terms of light absorption. Zhong $\mathit{et}$ $\mathit{al}$. uncovered a relationship between the tolerance factor~\cite{goldschmidt1926gesetze} and physical quantities such as band gap, dielectric constant, and Young's modulus, offering strategies for designing alloys for photovoltaic applications. Although Gebhardt $\mathit{et}$ $\mathit{al}$.~\cite{zhong2021structure} proposed hybrid antiperovskites, experimental investigations did not identify them as potential photovoltaic absorbers~\cite{gebhardt2017adding,gebhardt2018design}. However, both theoretical and experimental studies on various inorganic antiperovskite nitrides have revealed remarkable properties for photovoltaic applications~\cite{kang2022antiperovskite,kang2023native,dahbi2022earth,sreedevi2022antiperovskite,heinselman2019thin}.
	
To address related questions, we provide in this study a comprehensive picture of the electronic, optical, excitonic, and polaronic properties of Mg$_3$NSb, Sr$_3$NSb, Ba$_3$NSb, Ca$_3$NAs, and Sr$_3$NAs. Combining density-functional theory (DFT) with advanced hybrid functionals~\cite{krukau2006influence,heyd2003hybrid} and many-body perturbation theory (MBPT)~\cite{hedin1965new,hybertsen1985first}, we analyze the electronic properties, determine the electronic and ionic contributions to the dielectric screening, and compute exciton binding energies, electron-phonon coupling strengths, and charge-carrier mobilities. Calculations are performed with the PAW code VASP and the all-electron package \exciting. Computational details are provided in the Supplemental Material (SM).

\begin{figure*}
		\centering
		\includegraphics[width=0.8\textwidth]{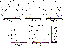}
		\caption{Atom-projected partial density of states (pDOS) of (a) $\mathrm{Mg_3NSb}$, (b) $\mathrm{Sr_3NSb}$, (c) $\mathrm{Ba_3NSb}$, (d) $\mathrm{Ca_3NAs}$, and (e) $\mathrm{Sr_3NAs}$ obtained by HSE06.}
	\label{dos}
\end{figure*}

To examine the structural stability, we estimate the Goldschmidt tolerance factor, $\mathit{t}$ ~\cite{goldschmidt1926gesetze} as detailed in Table 1 of the SM ~\cite{chi2002new,gabler2004sr3n,stoiber2019perovskite}. We find that Mg$_3$NSb and Sr$_3$NSb stabilize in cubic phases with tolerance factors between 0.9 and 1. In contrast, the other three materials, \ie Ba$_3$NSb, Ca$_3$NAs, and Sr$_3$NAs, possess distorted structures with smaller tolerance factors of 0.89, 0.86, and 0.85, respectively (see Section II of the SM), indicating increased stability of the distorted phases. Their dynamical stability is evidenced by the absence of negative frequencies in the phonon band structure, obtained by density functional perturbation theory (DFPT)~\cite{gajdovs2006linear} (Fig. S1 of the SM). 

Fig. ~\ref{dos} depicts the character of the states around the respective band gaps in terms of the atom- and symmetry-projected densities of states (pDOS) obtained by the HSE06 functional. In Mg$_3$NSb, the valence band maximum (VBM) is dominated by hybridization of N-$\mathit{p}$ and Sb-$\mathit{p}$ orbitals, while the conduction band minimum (CBM) is mainly composed of Mg-$\mathit{s}$ orbitals, enabling optical  transitions from $\mathit{p}$ to $\mathit{s}$ states. For the other materials, \ie Sr$_3$NSb, Ba$_3$NSb, Ca$_3$NAs, and Sr$_3$NAs, the VBM is mainly composed of hybridized  N-$\mathit{p}$ and A-$\mathit{p}$ orbitals, while X-$\mathit{d}$ orbitals largely contribute to the CBM. Therefore, dipole-allowed $\mathit{p}$ to $\mathit{d}$ transition are expected in these cases. We note that, in contrast to conventional halide perovskites, the A-site anion not only plays a role in the stability of these materials, but also makes a significant contribution to the band edges. The effective masses m$_{e/h}^*$ (see Table~\ref{bandgap}) extracted from the electronic band structures (see Fig. S2 of the SM) are found to be smaller than the free electron mass, indicating high charge-carrier mobilities and hence, good charge transport. Note that in Sr$_3$NSb, Ba$_3$NSb, and Ca$_3$NAs, m$_h^*$ $<$ m$_e^*$ \ie lighter holes at the valence band edges and thus, ultimately, high hole mobilities. These effective masses are in close agreement with previously reported values~\cite{mochizuki2020theoretical, kang2022antiperovskite}.
        
Band-structure calculations (Fig. S2) furthermore reveal that all studied nitrides except Mg$_3$NSb are direct-band-gap semiconductors with the VBM and CBM at the $\Gamma$ point, while in Mg$_3$NSb, the CBM is found at the M point. HSE06 (see Table~\ref{bandgap}) reproduces experimental ~\cite{heinselman2019thin,gabler2004sr3n} and previously reported theoretical values~\cite{mochizuki2020theoretical}. $\mathit{G_0W_0}$ based on HSE further opens all gaps. The values of 1.41 eV (Mg$_3$NSb),  1.14 eV (Sr$_3$NSb),  1.57 eV (Ba$_3$NSb), 1.91 eV (Ca$_3$NAs), and 1.52 eV (Sr$_3$NAs) all being below 2 eV, make them suitable for photovoltaic applications. The results obtained for Sr$_3$NSb is in excellent agreement with experiment~\cite{gabler2004sr3n}, $\mathit{G_0W_0}$, however, overshoots for Mg$_3$NSb \cite{heinselman2019thin}. 

	\begin{table*}
		\begin{center}
			\caption{Effective electron (m$_e^*$) and hole (m$_h^*$)  masses, and reduced masses ($\mu$) in units of the electron rest mass m$_0$ and fundamental band gaps (in eV) for the antiperovskites X$_3$NA, computed with different methodologies including SOC. Values obtained by \exciting\ are given in parentheses. With the exception of Mg$_3$NSb, all gaps are direct gaps at the $\Gamma$ point.}
   \vspace{0.2cm}
			\label{bandgap}
			\begin{tabular}{ c | c | c | c | c | c | c | c | c }
      \textbf{Material}  & \textbf{$\mathbf{m_e^*}$} & \textbf{$\mathbf{m_h^*}$} &\textbf{$\mathbf{\mu}$} & \multicolumn{5}{c}{{\bf Band gap}} \\
      \hline
				   & \multicolumn{3}{c|}{{\bf HSE06}} & \textbf{PBE} & \textbf{HSE06} & \textbf{$\mathit{G_0W_0}$$\mathbf{@PBE}$} & \textbf{$\mathit{G_0W_0}$$\mathbf{@HSE06}$} & \textbf{Experiment} \\
				\hline
				Mg$_3$NSb & 0.18 & 0.34 &  0.12  & 0.55 (0.54) & 1.23 (1.18) & 0.88 & 1.41 & 1.30$^d$, 1.10$^i$~\cite{heinselman2019thin} \\
				\hline
				Sr$_3$NSb & 0.23 & 0.21 & 0.11  & 0.16 (0.19) & 0.87 (0.87) & 0.52 & 1.14  & 1.15~\cite{gabler2004sr3n} \\
				\hline
				Ba$_3$NSb & 0.76 & 0.17 & 0.14  & 0.50 & 1.16 & 1.36 & 1.57 & - \\
				\hline
				Ca$_3$NAs & 0.37 & 0.32 & 0.17  & 0.82 & 1.63 & 1.64 & 1.91 & - \\
				\hline
				Sr$_3$NAs & 0.14 & 0.62 & 0.12 & 0.71 & 1.56 & 1.14 & 1.52 & - \\
			\end{tabular}
		\end{center}
	\end{table*} 

While these materials exhibit favorable band gaps and low effective carrier masses, these attributes alone do not guarantee their suitability as photovoltaic absorbers. In particular, the formation of excitons, \ie bound electron-hole pairs, may impact their transport properties. Therefore, we explore their optical response and excitonic properties at the absorption onset.

	\begin{figure*}
		\centering
		\includegraphics[width=1\textwidth]{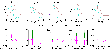}
		\caption{Real and imaginary part of the dielectric function (top) of the investigated antiperovskites obtained by BSE /$\mathit{G_0W_0}$. The ionic contributions to the dielectric screening are shown in the bottom panels.}
		\label{Fig3}
	\end{figure*}
	
For the lowest optical excitation, the exciton binding energy $E_{b}$ can be defined as the difference between the fundamental and the optical gap. Small exciton binding energies, as observed in these compounds, are typically confined in $\mathbf{k}$ space, \ie transitions taking place in a very small region around the band extrema. This poses significant computational challenges for the $\mathit{k}$-grid convergence of Bethe-Salpeter equation (BSE) calculations \cite{salpeter1951relativistic,rohlfing1998electron,albrecht1998ab}, in particular on top of $\mathit{G_0W_0}$$\mathrm{@HSE06}$ (see Section VI of the SM). Therefore, we use the Wannier-Mott model to estimate the exciton binding energies. According to this model, $E_b$ is expressed as
\begin{equation}
E_b = \frac{\mu}{\epsilon_{eff}^2} R_\infty ,
\label{eq:E_b}
\end{equation}
where $\mu$ is the reduced mass, R$_\infty$ is the Rydberg's energy, and $\epsilon_{eff}$ is the effective dielectric constant which lies between the high-frequency value $\epsilon_\infty$ and the long-wavelength limit $\epsilon_0$. To the best of our knowledge, no experimental values for $\epsilon_{eff}$ have been reported in the literature. In non-polar materials, typically $\epsilon_0$ does not differ from $\epsilon_\infty$. In other words, contributions from the lattice screening can be ignored. This is, however, not the case in polar materials like perovskites ~\cite{bechstedt2019influence,schleife2018optical,mahanti1970effective,mahanti1972effective}. Depending on to which extent, the lattice contributions are effective, one ends up with an upper and a lower bound to the binding energy (Table ~\ref{ebul}). 

\begin{table}
\begin{center}
\caption{Upper bound ($E_{b,max}$) and lower bound ($E_{b,min}$) of exciton binding energies (in meV), obtained by considering $\mathbf{\epsilon_{\infty}}$ or $\mathbf{\epsilon_0}$, respectively, in the Wannier-Mott model; LO phonon frequency ($\omega_{LO}$) at $\Gamma$; correction to the exciton binding energy by phonon screening (in meV), estimated using the static electronic dielectric constant ($\Delta E_b$), and contribution of this correction ($\frac{\Delta E_b}{E_b}\times100(\%)$).}
\label{ebul}
\begin{tabular}{ l | c | c | c | c | c | c | c }
\textbf{Material} & \textbf{$\mathbf{\epsilon_{\infty}}$} & \textbf{$E_{b,max}$}  & \textbf{$\mathbf{\epsilon_0}$} & $E_{b,min}$ & \textbf{$\mathrm{\omega_{LO}}$} & \textbf{$\mathrm{\Delta E_b}$} & $\frac{\Delta E_b}{E_b} (\%)$ \\
\hline
Mg$_3$NSb & 10.1 & 16 & 25.9  & 2 & 37 & -10  & 41  \\
\hline
Sr$_3$NSb & 7.0 & 31 & 28.9 & 2 & 27 & -13 & 39 \\
\hline
Ba$_3$NSb & 6.6 & 44 & 29.4 & 2 & 27 & -15 & 32 \\
\hline
Ca$_3$NAs & 4.2 & 134 & 25.6 & 4 & 36 & -26 & 19 \\
\hline
Sr$_3$NAs & 3.6 & 123 & 19.9 & 4 & 30 & -21 & 16 \\
\end{tabular}
\end{center}
\end{table}

In a recent study, Filip and colleagues \cite{filip2021phonon} included the phonon screening into the expression of $E_b$ using four distinct materials parameters \ie the reduced effective mass ($\mu$), static high and low frequency dielectric constants, and the longitudinal optical (LO) phonon frequency ($\omega_{LO}$). The electronic and ionic contributions to the dielectric screening are illustrated in Fig.~\ref{Fig3}. We have determined the characteristic frequency $\omega_{LO}$ from the multiple phonon branches using the athermal \lq B'\ scheme proposed by Hellwarth $\mathit{et}$ $\mathit{al}$. ~\cite{hellwarth1999mobility} (see Section VII of the SM for more details). The phonon-screening correction, assuming isotropic and parabolic electronic band dispersion, is given by  
\begin{equation}
    \Delta E_b= -2 \omega_{LO} \bigg(1 - \frac{\epsilon_\infty}{\epsilon_0}\bigg) \frac{\sqrt{1 + \frac{\omega_{LO}}{E_b}} + 3}{\bigg(1 + \sqrt{1 + \frac{\omega_{LO}}{E_b}}\bigg)^3}
    \label{eq:delE_b}
\end{equation}
Using this expression, we find (Table \ref{ebul}) that the contribution of phonon screening to the exciton binding energy is significant in these materials and hence, cannot be neglected. After incorporating the phonon-screening correction, the exciton binding energies are 6 meV, 18 meV, 29 meV, 108 meV, and 102 meV for Mg$_3$NSb, Sr$_3$NSb, Ba$_3$NSb, Ca$_3$NAs, and Sr$_3$NAs, respectively. Notably, for Mg$_3$NSb and Sr$_3$NSb, $E_b$ is smaller than k$_B$T at room temperature, where k$_B$ is the Boltzmann constant, while for Ba$_3$NSb, it is slightly larger. This implies that excitons can easily dissociate into free charge carriers in these materials. However, Ca$_3$NAs and Sr$_3$NAs show higher exciton binding energies. This behavior is also reflected in the exciton temperature, \ie the temperature up to which an exciton is stable before dissociating into a free electron and a free hole. It can be estimated as ${T_{exc}=E_b/ k_B}$. The resulting values along with other excitonic properties are summarized in Table ~\ref{exparam}. 

We further determine the exciton radius, ${ r_{exc} = \frac{m_0}{\mu}\varepsilon_{eff} a_{0} }$, where ${ a_0=0.529 \AA }$ is the Bohr radius. We find that in all investigated materials, the electron-hole pairs are distributed over many lattice constants, justifying the use of the Wannier-Mott model. For good photo-conversion efficiency of solar cells, the exciton lifetime  $\tau_{exc}$ should be high enough so that the photo-generated charge carriers can be extracted before recombination. $\tau_{exc}$ is proportional to the inverse probability of the electron-hole wavefunction at zero separation, $\mathrm{|\phi_n(0)|^2}$, \ie ${ \tau_{exc} \propto 1/|\phi_n(0)|^2 }$ with ${|\phi_n(0)|^2 = 1/\pi r_{exc}^3}$. We observe the following trend: $\tau_{exc}$(Ca$_3$NAs) $<$ $\tau_{exc}$(Sr$_3$NAs) $<$ $\tau_{exc}$(Ba$_3$NSb) $<$ $\tau_{exc}$(Sr$_3$NSb) $<$ $\tau_{exc}$(Mg$_3$NSb). In summary, the antimonide nitrides X$_3$NSb (X = Mg, Sr, Ba) have lower exciton binding energies, larger exciton radii, and longer exciton lifetimes than Ca$_3$NAs and Sr$_3$NAs, owing to relatively larger dielectric screening of the Coulomb interaction between electrons and holes, thus indicating better performance in photovoltaic applications.

	\begin{table}
		\begin{center}
			\caption{Exciton parameters for the studied antiperovskites. }
            \label{exparam}
			\begin{tabular}{ c | c | c | c | c }
				\textbf{Material}& $\mathbf{E_b}$ (meV) & $\mathbf{T_{exc}}$ (K) & $\mathbf{r_{exc}}$ (nm) & $\mathbf{|\phi_n(0)|^2} (10^{24} m^{-3})$ \\
				\hline
				Mg$_3$NSb & 6 & 110 & 5.8 & 1.6 \\
				\hline
				Sr$_3$NSb & 18 & 216 & 4.3 & 3.9 \\
				\hline
				Ba$_3$NSb & 29 & 342 & 3.0 & 11.4 \\
				\hline
				Ca$_3$NAs & 108 & 1269 & 1.4 & 108.5 \\
				\hline
				Sr$_3$NAs & 102 & 1198 & 1.8 & 58.9 \\
			\end{tabular}
		\end{center}
	\end{table}

Polar materials often experience polaron formation, which affects the motion of charge carriers. This interaction of charge carriers with the polar lattice vibrations is well described by the Fr\"ohlich model \cite{frohlich1954electrons}. Fr\"ohlich suggested a dimensionless descriptor to quantify the strength of this interaction
	\begin{equation}
		\mathrm{\alpha = \frac{1}{\epsilon^*} \sqrt{\frac{R_\infty}{\hslash \omega_{LO}}} \sqrt{\frac{m^*}{m_0}}},
	\end{equation}
where $\omega_{LO}$ is the LO phonon frequency, $\hslash$ is the Planck constant, and ${\frac{1}{\epsilon^*} = \frac{1}{\epsilon_\infty} - \frac{1}{\epsilon_0}}$ is an effective screening parameter. 
The computed carrier-phonon coupling constants ($\alpha$), polaron masses ($\frac{m_p}{m^*}$), polaron radii ($r_f$), and carrier mobilities ($\mu$) are listed in Table~\ref{polaron}. For both conduction electrons and holes, we observe that $\alpha$ $\leq$ 2, indicating weak coupling of the charge carriers with phonons. In this weak coupling regime, large polarons are formed, whose sizes we can estimate by the Schultz polaron radius, $r_f$ \cite{schultz1959slow}, confirming that for all materials, the polaron radii extend over multiple unit cells. 

Feynman extended Fr\"ohlich's polaron model to give an expression for the effective polaron mass, ${m_p = m^* \left( 1 + \frac{\alpha}{6} + \frac{\alpha^2}{40} + ... \right)}$ \cite{feynman1955slow}. The increase of the effective electron and hole masses upon polaron formation can be quantified to be between 7 $\%$ and 43 $\%$ in these materials. We further estimate an upper bound of charge carrier mobilities ($\mu_{e/h}$) using Feynman's variational solution of Fröhlich’s polaron Hamiltonian and by integrating the polaron-response function \cite{osaka1959polaron,hellwarth1999mobility,feynman1955slow,Frost2017} (see Section VII of the SM). We find that Mg$_3$NSb, Sr$_3$NSb, and Ba$_3$NSb possess exceptionally large carrier mobilities as compared to conventional perovskites, owing to weak electron-phonon coupling and formation of large polarons \cite{sendner2016optical,jain2022lead,jain2021theoretical}. Note that this approach neglects polaron-induced lattice distortions and scattering processes due to the acoustic phonons \cite{peter2010fundamentals}. Therefore, the actual charge carrier mobility may be lower than $\mu$.
\begin{table}
	\begin{center}
		\caption{Polaron parameters corresponding to electrons (e) and holes (h) in the studied antiperovskites.}
		\label{polaron}
		\begin{tabular}{ c | c | c | c | c | c }
		\textbf{Material} & \textbf{Carrier} & $\mathbf{\alpha}$ & $\mathbf{m_{p}/m^*}$ & $\mathbf{r_f}($\AA$)$ & $\mathbf{\mu (cm^{-2}/Vs)}$ \\
		\hline
		Mg$_3$NSb & e & 0.4 & 1.07 & 186 & 713 \\
    	   & h & 0.6 & 1.11 & 156 & 323\\  
		\hline
		Sr$_3$NSb & e & 0.8 & 1.15 & 102 & 363 \\
            & h & 1.2 & 1.24 & 86 & 127\\
        \hline
		Ba$_3$NSb & e & 1.0 & 1.19 & 94 & 228 \\
          & h & 0.9 & 1.17 & 99 & 210 \\
		\hline
		Ca$_3$NAs & e & 1.6 & 1.33 & 90 & 98 \\
             & h & 1.4 & 1.28 & 98 & 64 \\
	      \hline
		Sr$_3$NAs & e & 1.7 & 1.36 & 77 & 135 \\
           & h & 2.0 & 1.43 & 70 & 21 \\
		\end{tabular}
		\end{center}
\end{table}

In summary, we have conducted a detailed analysis of electronic, optical, excitonic, and polaronic properties in inorganic antiperovskite nitrides. Our electronic-structure calculations reveal effective electron and hole masses smaller than the free electron mass, indicative of high charge-carrier mobilities. We find that all considered materials have band gaps in the visible range of the solar spectrum, making them suitable for photovoltaic applications. By employing the Wannier-Mott model, we obtain low exciton binding energies, specifically in Mg$_3$NSb, Sr$_3$NSb, and Ba$_3$NSb. This indicates that free charge carriers are predominantly generated upon light absorption. The computed Fr\"ohlich coupling constants indicate weak electron-phonon coupling and thus formation of large polarons in all studied antiperovskites. We find high charge-carrier mobilities in $\mathrm{Mg_3NSb}$, $\mathrm{Sr_3NSb}$, and $\mathrm{Ba_3NSb}$, as estimated by the Hellwarth polaron model. Moreover, despite exhibiting an indirect band gap, among the studied antiperovskites, $\mathrm{Mg_3NSb}$ has the highest mobilities of 713 $\mathrm{cm^{-2}/Vs}$ and 323 $\mathrm{cm^{-2}/Vs}$ for electrons and holes, respectively. From all these results we suggest $\mathrm{Mg_3NSb}$, $\mathrm{Sr_3NSb}$, and $\mathrm{Ba_3NSb}$ to be favorable candidates for photovoltaic applications.

S.M. acknowledge IIT Delhi for the junior research fellowship and the NOMAD Center of Excellence for funding the internship programme at the Humboldt-Universität zu Berlin, supported by the European Union’s Horizon 2020 research and innovation program under the grant agreement Nº 951786. M.J. acknowledge CSIR, India, for the senior research fellowship [Grant No. 09/086(1344)/2018-EMR-I]. S.B. acknowledge financial support from SERB under a core research grant [Grant no. CRG/2019/000647] to set up his High Performance Computing (HPC) facility ``Veena'' at IIT Delhi for computational resources. CD appreciates partial support from the SPP program 2196, project 24709454, funded by the German Research Foundation DFG.
	
%
\end{document}